\documentclass{aa} 
\usepackage{txfonts}
\usepackage{graphicx}% Include figure files
\usepackage{hyperref}% add hypertext capabilities
\usepackage{soul}
\usepackage{color}

\DeclareMathAlphabet\mathbfcal{OMS}{cmsy}{b}{n}
\DeclareMathAlphabet\mathcal{OMS}{cmsy}{n}{n}

\def\aap{{Astronomy and Astrophys.}}

\def\apj{{\it Astrophys. J.\ }}
\def\apjs{{\apj\ \it Suppl.\ }}         
\def\apjl{{\apj\ \it Lett.\ }}

\def\ds{\displaystyle}
\def\vA{v_\mathrm{A}}
\def\MA{M_\mathrm{A}}
\def\ne{n_\mathrm{e}}
\def\ni{n_\mathrm{i}}
\def\pe{p_\mathrm{e}}
\def\me{m_\mathrm{e}}
\def\mi{m_\mathrm{i}}

\def\v0e{v_\mathrm{0e}}
\def\omegap{\omega_\mathrm{p}}
\def\omegape{\omega_\mathrm{pe}}
\def\omegac{\omega_\mathrm{c}}
\def\omegace{\omega_\mathrm{ce}}
\def\Te{T_\mathrm{e}}
\def\Ti{T_\mathrm{i}}
\begin{document}

\title{Electron energization in quasi-parallel shocks}

   \subtitle{Test-particle electrons in a proton-driven turbulence}

   \author{Adrian Hanusch
          \inst{1}
           \and
          Tatyana V. Liseykina\inst{1,2,3}
          \and Mikhail A. Malkov\inst{4}
                    }

   \institute{Institut f\"ur Physik, Universit\"at Rostock, 18051 Rostock, Germany\\
              \email{adrian.hanusch@uni-rostock.de}
                      \and
         Institute of Computational Mathematics and Mathematical Geophysics SD RAS, 630090, Novosibirsk Russia
         \and
              Nikolsky Institute of Mathematics, RUDN, 115419 Moscow, Russia\\
             \email{tatyana.liseykina@uni-rostock.de}
                          \and
                CASS and Department of Physics, University of California, San Diego, La Jolla, California 92093, USA\\
             \email{mmalkov@physics.ucsd.edu}
                                                    }

\date{\today}% It is always \today, today,
             \abstract{
             %\LEt{ Only the first word of titles and subheadings should be capitalized. + I edited your paper to US\ spelling and grammar conventions.}
             In situ observations of energetic particles at the Earth's bow-shock that are attainable by the satellite missions have fostered the opinion for a long time that electrons are most efficiently accelerated in a quasi-perpendicular shock geometry.
%, while quasi-parallel shock geometry is more favorable for the acceleration of protons. 
 However, shocks that are deemed to be responsible for the production of cosmic ray electrons and their radiation from sources such as supernova remnants are much more powerful and larger than the Earth's bow-shock. Their remote observations and also 
 %\LEt{ Please avoid the use of italics for emphasis. Please check for this throughout the paper. See Sect. 1.2 of the Author's Guide.}
 in situ measurements at Saturn’s bow shock, that is, the strongest
shock in the Solar System, suggest that electrons are accelerated very efficiently in the quasi-parallel shocks as well.} {In this paper we investigate the possibility that protons that are accelerated to high energies create sufficient wave turbulence, which is necessary for the electron  preheating and subsequent injection into the diffusive shock acceleration  in a quasi-parallel shock geometry.} {
  An additional test-particle-electron population, which is meant to be a low-density addition to the electron core-distribution on which the hybrid simulation operates, is introduced. Our purpose is to investigate how these electrons are energized by the "hybrid" electromagnetic field. 
  The reduced spatial dimensionality allowed us to dramatically increase the number of macro-ions per numerical cell and achieve the converged results for the velocity distributions of test electrons.} {We discuss the electron preheating mechanisms, 
which can make a significant part of thermal electrons accessible to the ion-driven waves observed in hybrid  simulations. We find that the precursor wave field supplied by ions has a considerable potential to preheat the electrons before they are shocked at the subshock. Our results indicate that a downstream thermal equilibration of the hot test electrons and protons does not occur. Instead, the resulting electron-to-proton temperature ratio is a decreasing function of the shock Mach number, $M_\mathrm{A},$ which has a tendency for a saturation at high $M_\mathrm{A}.$} 
{}
%\pacs{98.38.Mz, 98.70.Sa}% PACS, the Physics and Astronomy
\keywords{cosmic rays, ISM:supernova remnants, acceleration of particles, methods:numerical}
\maketitle

\section{Introduction}
Collisionless shocks are ubiquitous in astrophysical environments and are
proven to be efficient particle accelerators, with supernova remnant 
(SNR) shocks being the most probable source of galactic cosmic rays 
\citep{Gaisser1991}.
The acceleration of charged particles at these shocks is accurately described 
by the theory of diffusive shock acceleration (DSA, for a recent review, see \citep{Schure2012}. While this mechanism
is conceptually simple, its precise outcome for the energy spectra and chemical composition of accelerated particles is difficult to determine under realistic shock conditions \citep[see e.g.,][]{Ohira16,Hanusch2019_steepening}.
In particular, the 
"injection" of different species into the DSA remains largely unsolved \citep{CaprioliPRL17,Hanusch2019,Evoli-PRD2019},
with the electron injection being a notoriously difficult problem \citep[see e.g.,][]{West-AA2017,Aharonian-AA2017}. 
An important question is whether protons that are accelerated to high energies
create the sufficient wave turbulence required for the electron injection into
the DSA. Electrons, in contrast to ions, have long been thought to be incapable of injecting themselves on self-generated waves, especially in quasi-parallel shocks, 
so some ion "assistance" appears critical.

The acceleration of electrons at quasi-perpendicular collisionless
shocks has been investigated by means of numerical simulations by a number of authors. \citet{Riquelme2011} performed fully kinetic simulations and have shown that whistler waves are crucial for electron injection.
%\LEt{ Please replace the semicolons with commas and add ", and" before Matsumoto.\ Unfortunately, I cannot do so on my end.} 
\citet{Shimada2000}, \citet{Hoshino2002}, \citet{Amano2008}, \citet{Matsumoto2012}, and \citet{Matsumoto2017} report that electrostatic waves excited by the Buneman instability and accompanied by the particles trapping are important for efficient electron acceleration. Even though \citet{Matsumoto2012} clarify electron-to-ion mass ratio dependence of Buneman instability, it is unclear, whether this mechanism is robust if a real electron-to-ion mass ratio is considered. Artificially enhanced electron-to-ion mass ratios, which are often used in fully kinetic simulations  in order to obtain a converged result within finite time, may lead to physically questionable results, as discussed, for example, in  
%\LEt{ Please replace the semicolon with "and".\ For issues of this nature that require your attention later on in the text, I'll leave the note "punctuation issue" .}
\cite{Matsukiyo2006}, \cite{Bohdan2019-1}, and \cite{Bohdan2019-2}. 
An alternative approach was chosen by \cite{Burgess2006}, \cite{Guo2010}, and \cite{Trotta2018}.  
These authors investigated the effect of the shock surface and magnetic field fluctuations
on electron acceleration by following the trajectories of test-particle electrons in the fields obtained from hybrid simulations.
Kinetic simulations that were performed to investigate the injection of electrons
at quasi-parallel shocks \citep[e.g.,][]{Park2015},
have revealed the  injection into the DSA  with the scattering of both ions and electrons.  Waves that are excited via Bell instability \citep{Bell2004} mediate the scattering process. 
Yet another method  
was chosen by \cite{Guo2015}, where the authors followed the trajectories of the test electrons in the prescribed, kinematically-defined electromagnetic fields in the shock region and they studied the dependence of the electron acceleration efficiency on the shock inclination and wave variance. In particular, 
\cite{Guo2015} have shown that the acceleration of these particles does not strongly depend on the shock inclination unless the upstream turbulent magnetic field is weak.
%\st{only strongly depends on the shock inclination if the upstream magnetic field turbulence is not sufficiently large.} 
On the contrary, the electrons are found to get efficiently accelerated at quasi-parallel shocks as well, provided the upstream magnetic field fluctuations are strong enough. This finding is consistent with NASA's earlier Cassini spacecraft observation of Saturn's bow shock \citep{Masters2013}, which for the first time provided in situ evidence of a significant electron acceleration in a quasi-parallel shock geometry\footnote{We note that for the quasi-parallel spacecraft crossing, the strong evidence for shock acceleration of electrons was only found at
$\MA=100.$ 
This suggested that the electron acceleration resulted from the unusually high $\MA$.}. The kinetic model of electron injection and acceleration at quasi-parallel supercritical collisionless shocks, which was developed by  \cite{Bykov1999},  demonstrated that strong MHD fluctuations generated by ion kinetic instabilities are important for heating and the pre-acceleration of suprathermal electrons on very small scales.

For the analysis 
of observed X-ray spectra as well as in order to understand the energy partitioning 
between the energetic cosmic-ray (CR) and the thermal populations, electron acceleration is important along with electron thermalization.  As the transition between the unshocked and shocked medium is much shorter than the collisional mean free path, an equilibration of ion and electron temperatures may only occur on long timescales. Observations of Balmer-dominated shocks have shown a dependence of the electron-to-ion tem\-pe\-ra\-tu\-re ratio on the shock velocity \citep{Ghavamian2013}. Fully kinetic particle-in-cell simulations of low Mach number quasi-perpendicular shocks also indicate that the electron-to-proton temperature ratio of the shocked medium is a function of the shock Mach number \citep{Guo2018}.

    In this paper,  we study the electron-ion temperature relaxation in a quasi-parallel shock geometry by introducing electrons as test particles in
    hybrid simulations and we investigate their thermalization in the proton-driven 
    turbulence. We compute the trajectories of these electrons, but we assume that their contribution to 
    %\LEt{ Please check the intended meaning hasn't changed.}
    charges and currents is negligible, so the electromagnetic fields are not affected.
 It should be emphasized that 
although hybrid simulations neglect the contribution of suprathermal electrons, assuming that their number is small enough so as to not affect the simulation results significantly, this low-density electron component, which is negligible for the simulations dynamics, is very important observationally.
On the other hand, if hybrid modeling provides realistic field distributions,
%\st{ which have relevance to  the reality},
%\LEt{ Please specify what you mean here, "the reality of...".}
and there exist reasons to believe this, the study of the electron behavior in these fields is also worthwhile. However limited, this approach elucidates aspects of electron heating that are not accessible to hybrid simulations.

Our choice of a one-dimensional (1D) simulation is 
motivated by computational economy, spatial resolution reasoning, and physical considerations.
%\st{not only motivated by computational economy and spatial resolution reasoning, but also by physical considerations.} 
First of all, a realistic alternative to 1D simulations would be a two-dimensional (2D)  simulation. However, the 2D simulations also have drawbacks that we have recently discussed \citep{Hanusch2019}. We argue that when the  high resolution 
and particle statistics are the priority, a 1D code may be a better choice. Of course, when the 2D effects are crucial, as in the case of acceleration at the variable shock obliquity \citep{Hanusch2019_steepening}, the 1D simulation setup is not a possibility.  

The problems associated with the 2D option are primarily due to inverse turbulent cascades, which are not present in the real three-dimensional (3D) systems. Additionally, and more importantly for this study, the conserved particle canonical momentum component in the direction of the ignorable coordinate, in combination with strong magnetic eddies produced by the inverse cascade, result in a protracted interaction of particles with the eddies.
This interaction is akin to the shock drift (or the so-called surfatron) acceleration, occurring when a particle "surfs" on the edge of an eddy. It can strongly modify the particle transport, both in momentum and coordinate space. This phenomenon has recently been studied by a direct comparison of 2D and 3D simulations by \cite{Trotta2018}.

\section{Electron preheating mechanisms}
The investigation of electron energization by the electromagnetic fields generated
in hybrid simulations implies the existence of independent preheating
mechanisms for this electron population. 
%Indeed, the hybrid simulations
%treat electrons as a fluid so that they can interact with relatively
%long waves that are generated by nonequilibrium ion populations (such as shock-reflected
%ions) within the hybrid simulations 
%\LEt{ I suggest rewording here and possibly writing "they can only interact adiabatically" instead.}
%only adiabatically. 
Indeed, the hybrid simulations
treat electrons as a fluid so that their interaction with relatively
long waves that are generated by nonequilibrium ion populations (such as shock-reflected
ions) within the hybrid simulations can only occur adiabatically.
These waves
cannot heat the electrons appreciably. However, these are not the
only waves that are generated by these types of ions in the real shocks. Much
shorter waves with higher frequencies, which are not accessible to hybrid simulations,
may also be generated. They can tap into thermal electrons and preheat
them. Then, electrons start to interact with the waves generated in
the hybrid simulations. 

\begin{figure}[t]
\includegraphics[width=0.97\linewidth]{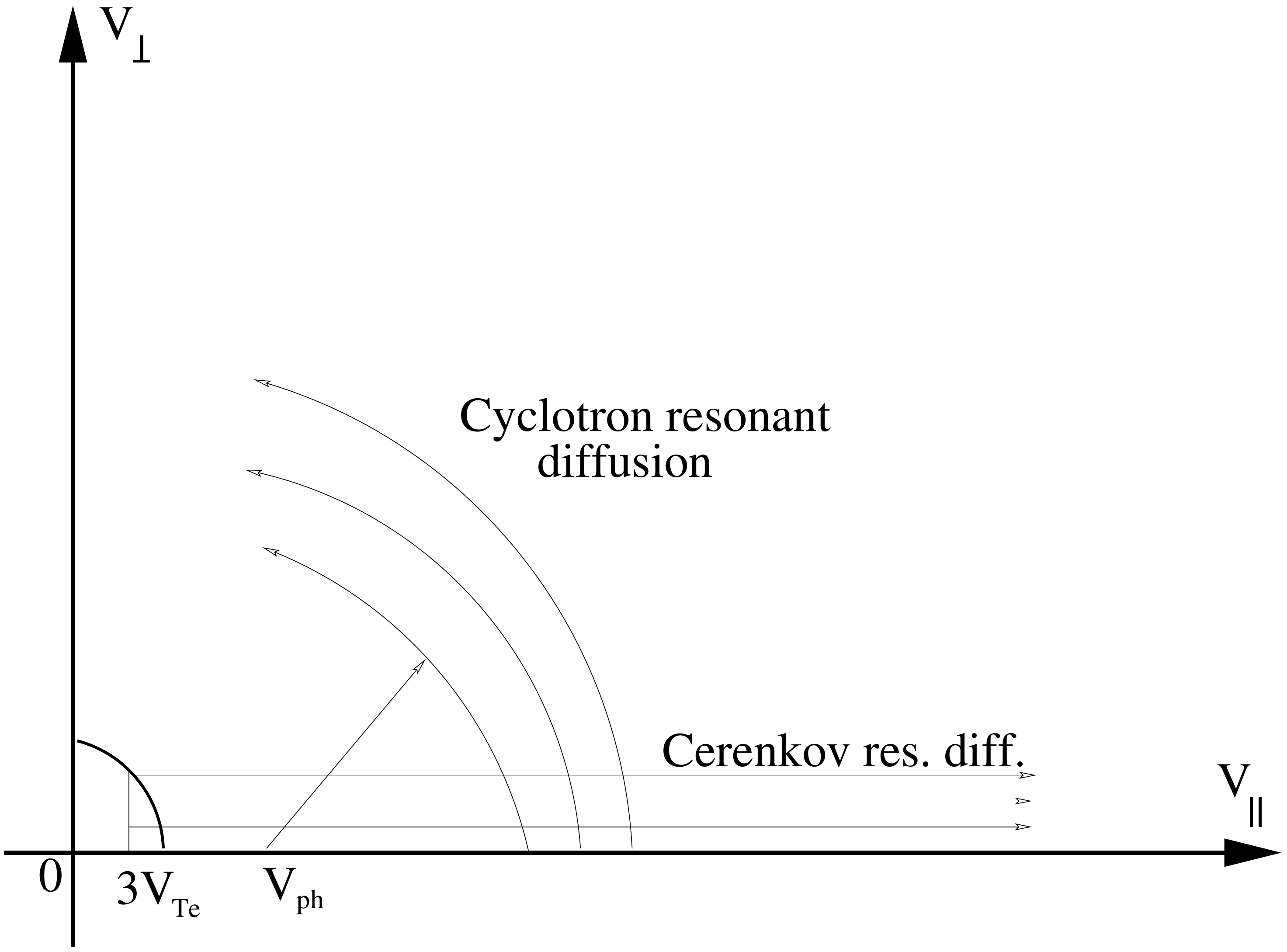}
\caption{Electron preheating in high frequency plasma waves, primarily the
lower-hybrid and oblique Langmuir waves, $V_{\mathrm{ph}}=\omega/k_{\parallel}\simeq\omegace/k$,
that are not accessible to the hybrid simulations.\label{fig:Electron-preheating-in}}
\end{figure}

The preheating mechanisms are not straightforward, and their thorough
description is outside the scope of this short paper, while a brief
overview is in order. They have been considered in many earlier publications,
starting perhaps from those in the magnetic fusion research and general
plasma physics, %\LEt{ Please add "e.g.," inside the parenthesis.}e.g.,
\citep[e.g.,][]{Shapiro1968JETP}. These types of mechanisms invoke, almost universally, a combination of Cerenkov and cyclotron resonances
impacting the same particle populations. The wave modes associated
with these resonances may or may not be the same. The salient aspects
of the wave-particle interaction are briefly explained in Fig.\ref{fig:Electron-preheating-in}. 

Suppose that the initial population of electrons is stable, for example,
a Maxwellian, but there are waves accelerating electrons in the velocity
component in the magnetic field direction, $v_{\parallel}$, by a
quasilinear diffusion. The lower-hybrid waves are, perhaps, most potent.
They have an approximately constant frequency, $\omega\approx\sqrt{\omegac\omegace}$
(in the dense plasma limit $\omegac\omegace/\omegap^2=\omegace^2/\omegape^2 \ll 1$).
Here $\omegace$ and  $\omegape$ are the electron cyclotron and plasma frequencies, respectively. 
When propagating at large angles to the local magnetic field ($k_{\perp}\gg k_{\parallel}$),
they accelerate electrons over a broad range in $v_{\parallel}$
via a Cerenkov resonance, $\omega-k_{\parallel}v_{\parallel}=0$.
However, these waves are damped by the same electrons and cannot tap
into their distribution deeper than, typically, $v_{\parallel}\simeq3V_{T_\mathrm{e}}$
because $\left|\partial f_\mathrm{e}/\partial v_{\parallel}\right|$ increases
at lower $v_{\parallel}$ and leads to strong wave damping. Energy-wise,
however, these waves, which are driven by powerful ion populations which also happen to be observed in hybrid simulations, could pull more electrons
out of the core Maxwellian. They are just not excited for resonant
velocities of $v_{\parallel}=\omega/k_{\parallel}<3V_{T_\mathrm{e}}.$ %\st{, for example.} 

Nevertheless, there are still mechanisms whereby the low-energy
electrons with $v_{\parallel}\lesssim3V_{T_\mathrm{e}}$ can also be accelerated.
One such mechanism has been discussed in conjunction with the electron
injection into the DSA \citep{GMV95}. By
this mechanism, a macroscopic 
%\st{microscopic} 
electric field is generated in response
to the acceleration of an initially small fraction of electrons by
the lower-hybrid waves in the region of $v_{\parallel}>3V_{T_\mathrm{e}}$, for instance. While
these electrons tend to escape the spatial region of their acceleration,
the emerging charge imbalance must be neutralized by the electric
field in the shock precursor. This field also accelerates electrons
with lower energies, placing them into the region of $v_{\parallel}>3V_{T_\mathrm{e}}$
where they, being picked up by the waves, add up to the 
%\LEt{ "already preheated" is redundant.\ I suggest writing just "preheated" or "already heated".}already 
already heated
electron population. 

The second mechanism is also based on the electron distribution that is already
stretched along $v_{\parallel}$. It may easily become unstable with
respect to the wave generation via cyclotron resonance, $\omega-k_{\parallel}v_{\parallel}+n\omegace=0,$ where
$n$ is an integer \citep{Shapiro1968JETP}. A quasilinear diffusion
then follows along the lines on the $v_{\parallel},v_{\perp}$ plane
that are determined by the relation $v_{\perp}^{2}+v_{\parallel}^{2}-2\int\left(\omega/k_{\parallel}\right)dv_{\parallel}=const$.
Here, the wave phase velocity $V_{\mathrm{ph}}=\omega/k_{\parallel}$
along the field should be expressed using the above resonance relation.
It may be seen from Fig.\ref{fig:Electron-preheating-in} that this
diffusion leads to energy losses of electrons in the wave frame, thus
making the wave unstable. An important aspect of this diffusion is
that electrons are swept up to lower $v_{\parallel}$ and higher $v_{\perp}$,
so that their distribution that is integrated over $v_{\perp}$ no longer
strongly contributes to the wave damping on the Cerenkov resonance
in the critical region $v_{\parallel}\simeq3V_{T_e}.$ As a
result, more thermal electrons fall into the Cerenkov resonance interaction
with the lower-hybrid waves and they get accelerated. 
To conclude this section, a significant part of thermal electrons
can be made accessible to the ion-driven waves observed in hybrid
simulations. Having discussed it as a proof of principle, we stop
short of making specific predictions about the exact number of these types of
electrons for two reasons. First, some useful information can be found
in the cited papers and references therein. Nevertheless, to our knowledge,
there are no such calculations broadly applicable to shocks. Otherwise,
a comprehensive theory of electron injection into the DSA would have
been developed. Second, a test-particle treatment of energetic electrons
in this paper places limits on their number, as is discussed in the
Introduction.

\section{Model and simulation setup \label{Sec:Model}}
The investigation of particle dynamics at collisionless shocks, in the context of CR acceleration at SNR shocks, has largely relied on numerical simulations. 
Although the in situ measurements made by spacecrafts can potentially shed light on the physics of these shocks, they are only attainable for shocks in the Solar System, such as Earth's bow shock (by satellite missions) \citep[e.g.,][]{Sundberg2016,Feldman1983,Amano2020} or Saturn's bow shock \citep{Masters2013}. Moreover, as Solar System shocks are much less powerful and smaller than SNR shocks, it is not clear whether the  dependence of the acceleration efficiency on the magnetic field orientation with respect to the shock normal, which is observed for the electrons in Solar-wind plasma shocks,  applies to SNR shocks as well.

While the use of fully kinetic simulations is the most fundamental approach, it is computationally very expensive and, in multidimensional geometry, it can even be unfeasible to follow the evolution of a collisionless shock over many ion cyclotron times. 
 In order to overcome these difficulties, unrealistic and strongly increased electron-to-ion mass ratios
are often used in kinetic simulations.

When focusing on the acceleration of ions, the hybrid approach has been proven to be a valuable tool \citep{Lipatov2013}. In these simulations, the electrons are treated as a charge neutralizing 
fluid. If moreover, one neglects the 
electron mass, the equation of motion of the 
electron fluid reduces to
\begin{equation}
    0 
    = -e\, \ne  \left( \vec{E} + \frac{1}{c} \, \vec{v}_\mathrm{e} \times \vec{B} \right) 
      - \nabla \pe + e \, \ne \, \eta\, \vec{J},
      \label{eq:e-fluid}
\end{equation}
where $-e$, 
$\ne$, and $\vec{v}_\mathrm{e}$ are the electron charge, 
density, and bulk velocity, respectively, and $\vec{J}$ is the total current. The last term on the right-hand side of \eqref{eq:e-fluid} describes the 
resistive coupling between electrons and ions.  A phenomenological anomalous resistivity $\eta$ gives rise to electron Ohmic heating and smooths the fields on the resistive scale-length. Both the resistivity  and pressure, $\pe$, are assumed to be scalar and an adiabatic equation of state with an adiabatic index of $\ds \gamma = 5/3$ is used for the fluid electrons.
In the hybrid model, the ions are treated kinetically, and their motion is governed by the following nonrelativistic equations
\begin{equation}
    \mi \, \frac{d \vec{v}}{dt} = q_\mathrm{i} \left( \vec{E} + \frac{1}{c} \, \vec{v} \times \vec{B} - \eta\, \vec{J} \right), \qquad
    \frac{d \vec{x}}{dt} = \vec{v}.
\end{equation}
In the simulations, lengths are normalized to the ion skin depth, $c/\omegap$, with the proton plasma frequency $\ds \omegap=\sqrt{4\pi\,n_{0}\,e^{2}/m_\mathrm{p}}$. Here, $n_{0}$ denotes the plasma density that is far upstream  and $e$ and $m_\mathrm{p}$ are the proton charge and mass, respectively. Time is measured in units of inverse proton gyrofrequency, $\ds \omegac^{-1}=(e\,B_{0}/m_\mathrm{p}\,c)^{-1}$, and velocity is measured in units of the Alfv\'en velocity, $\ds \vA = B_0 / \sqrt{4\pi\, n_0\, m_\mathrm{p}}$. Here, $B_{0}$ denotes the magnitude of the background magnetic field, which is set to $\vec B_0 = B_0 \, \vec e_x$, which is parallel to the shock normal ($x$-axis in  our  convention).  
\begin{figure}
    \centering
    \includegraphics[width=0.9\linewidth]{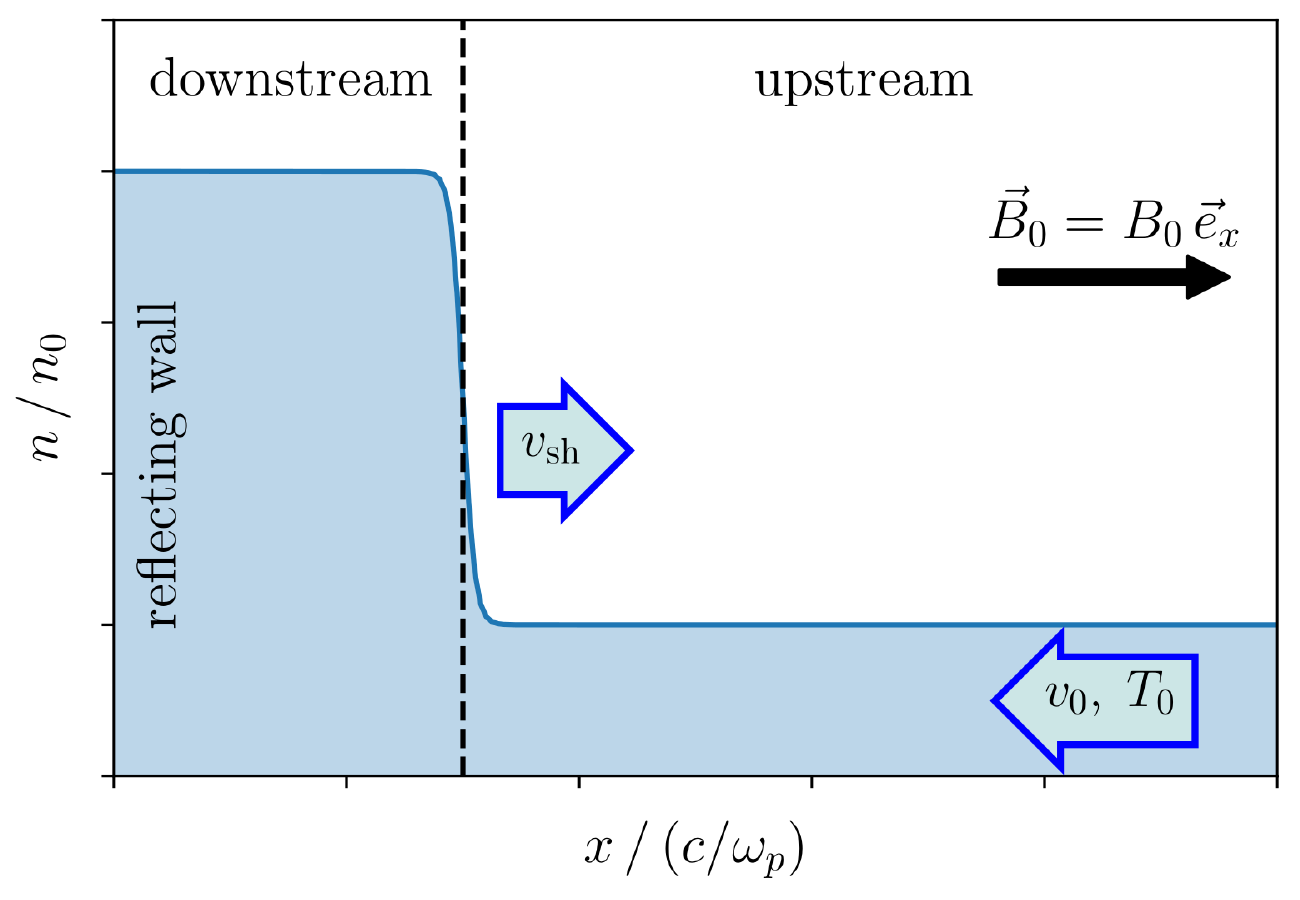}
        \caption{Simulation setup: a shock is created by
sending a super-sonic plasma flow with a velocity $v_0$ against a reflecting wall.
The shock propagates to the right, parallel to the background magnetic
field.}
        \label{fig:setup}
\end{figure}

In the following, we investigate the 
electron kinetics in the fields generated by the ions, which is beyond Eq.~\eqref{eq:e-fluid}.   If stochastic fields generated in hybrid simulations are realistic, electron orbits in these fields also deserve attention. To this end, we add a population of electrons as test particles in our simulation.  These electrons, by definition, do not generate electric or magnetic fields, nor do they exert pressure on the background plasma. 
 The idea is in the spirit of earlier simulations of quasi-perpendicular shocks, where the fields obtained from hybrid simulations were used to propagate test-particle electrons \citep{Burgess2006,Guo2010,Trotta2018}.  
 
Due to the separation of scales, we have introduced a sub-cycling routine in our hybrid numerical code to properly resolve the trajectories of the test-particle electrons. The propagation of the electrons is performed in $N_\mathrm{cyc}$ sub steps, reducing the effective time step for the electron propagation to $\Delta t_e = \Delta t / N_\mathrm{cyc}$. A linear interpolation between the fields known at the time-steps of the ion propagation $t = n\, \Delta t$ and $t = (n+1) \, \Delta t$ is used to obtain the fields at the sub steps. To reduce the numerical costs, we use a moderately  increased electron-to-proton mass-ratio of $\me/m_\mathrm{p} = 1/400$ and update the electron positions and velocities $N_\mathrm{cyc}=20$ times during one propagation step $\Delta t$ of the ions.
We note that the use of the guiding center approximation or a  gyro-kinetic treatment \citep{Frieman1982, Littlejohn1983} of the electrons would also be possible; however, in starting from some energy, adiabaticity can stop working well for the electrons.
The electron fluid is initially assumed to be in thermal equilibrium
with the ions with $\beta_\mathrm{e} = \beta_\mathrm{p} = 1$. The simulation is initialized by sending a super-sonic and superalfv\'enic  hydrogen plasma flow with velocity $v_0$ against a reflecting wall, placed  at $x=0,$ as is seen in Fig.~\ref{fig:setup}. A shock forms upon the interaction of the counter-propagating plasma streams and propagates in the positive $x$-direction. 
Since we are not sure about the heating mechanism, we considered different far-upstream distributions for the test-particle electrons as follows:
(T)  -- the mean velocity of the test-electron population equals the plasma flow upstream speed of $\vec{v}_\mathrm{0e}=\vec{v}_0$ and its temperature equals the temperature of the electron core distribution on which the hybrid simulation operates;
(B) "beam" -- the mean velocity of the test-electron population is significantly bigger than the far upstream plasma flow speed of $v_\mathrm{0e} \gg v_0$ and 
$\vec{v}_\mathrm{0e}\parallel\vec{v}_0,$ and its temperature equals the temperature of the electron fluid; (S) "shell" -- the energy of test electrons is significantly higher than the directional energy of the electrons in the plasma flow $v_\mathrm{0e}=100~v_A~\simeq 5 V_{\Te} \gg v_0$. The test-particle electrons are injected at the right boundary starting from $t=0$ (T) and  
from $t\omegac=50$ for
$\v0e/\vA=30$ and $\v0e/\vA=100,$ (B), respectively, 
with a velocity distribution according to a Maxwellian flux with a drift velocity of $\v0e$ (see \cite{Cartwright2000} for details). 
The simulations were performed with a temporal resolution of $\Delta t = 0.01\left(c/\omegap\right)/v_0$ and a cell size of $\Delta x = 0.25\, c/\omegap. $
At least 1000 ions per cell were used to keep the numerical noise\footnote{We have checked that when the number of ions per cell is smaller than $N_\mathrm{ppc}=400$, the numerical noise leads to an artificial electron heating and consequently to an electron velocity distribution that is strongly dependent on $N_\mathrm{ppc}$.} in the electromagnetic hybrid fields at a low level. 
To strongly improve particle statistics and avoid problems inherent in 2D simulations, as is briefly discussed in the Introduction, only one spatial dimension ($x$ in our convention), but all of the components of the velocity and fields, are included.
\section{Results}
We have performed hybrid simulations for different initial upstream flow velocities, giving rise to the formation of the shocks with 
different Mach numbers, and we followed the evolution of the shock for several hundreds of ion cyclotron times. 
In the following, we present the results of the simulations investigating the  behavior of test-particle electrons moving in the turbulent fields that are created by the ion plasma component. 

\begin{figure}
   \includegraphics[width=\linewidth]{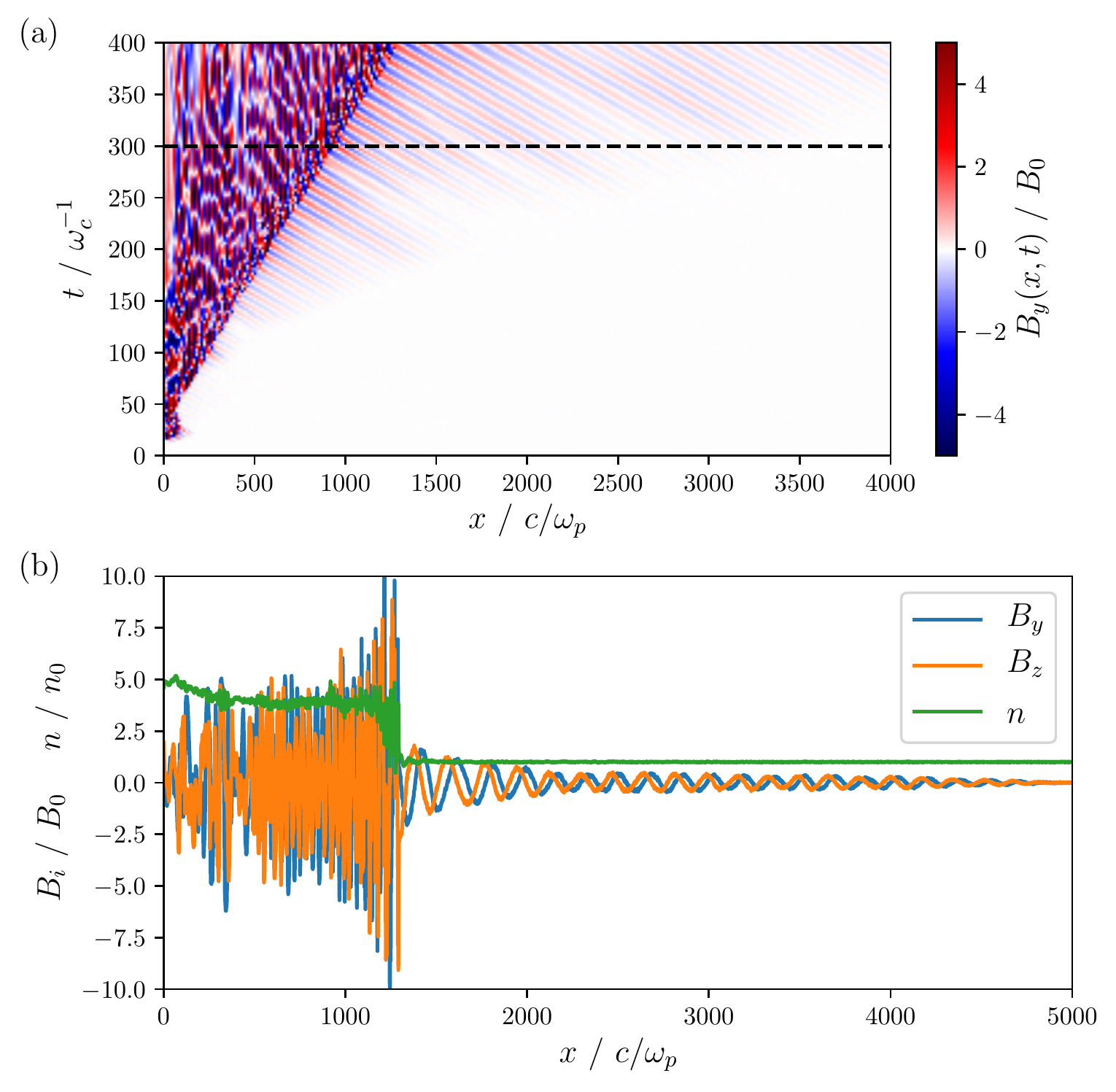}
        \caption{(a) Space-time plot of the $y$-component of the magnetic field.
                         (b) Ion density $\ni$ and the components of the magnetic field, $B_y$ and $ B_z$, for a
                         simulation with an upstream flow velocity of $\ds v_0/\vA = 10$ at 
                         $t\,\omegac = 300$.
                         }
        \label{fig:fields}
\end{figure}

Figure~\ref{fig:fields}a) shows the temporal evolution of the spatially dependent $B_y$ for a simulation with $v_0/\vA = 10.$ The shock propagates to the right and the compression of the magnetic field is clearly visible. The dashed line denotes the time at which the density  and components of the magnetic fields are plotted in Fig.~\ref{fig:fields}b).
It can be seen that the ion density increases upon the shock crossing. Circularly polarized Alfv\'en waves are excited by the streaming protons. As the magnetic field of these waves is almost frozen into the plasma ($\vA\ll v_0$), they are advected downstream and compressed, leading to large amplitudes of the magnetic field behind the shock front.

\begin{figure}[t]
    \includegraphics[width=\linewidth]{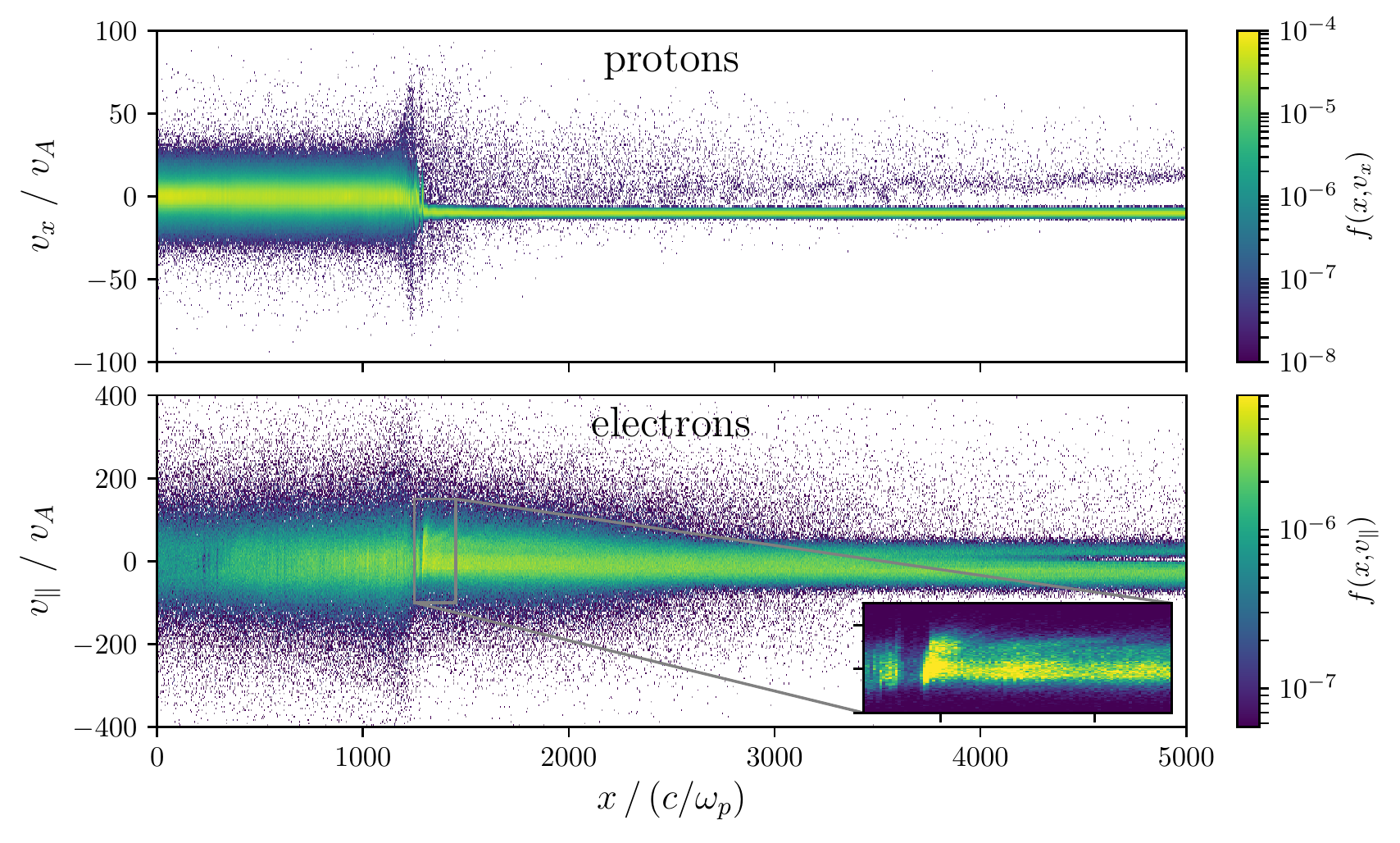}
        \caption{Phase-space $f_p(x, v_x)$ of protons (top) and  $f_e(x, v_\parallel)$ of 
                         test-particle electrons (bottom) at $\ds t \,\omegac= 400.$ The inset shows $f_\mathrm{e}(x, v_\parallel)$ to be close to the shock transition on a linear scale for a simulation (T) with $\ds v_0/\vA=10.$}
        \label{fig:phase_space}
\end{figure}

Detailed information about the accelerated particles and their temperature can be extracted from the particle phase spaces. Figure~\ref{fig:phase_space} shows the distributions of protons (top) on the $(x,v_x)$ plane and of the test electron (bottom) on the $(x,v_\parallel)$ plane  
at $t \,\omegac= 400.$  Here, $v_\parallel$ is the electron velocity component parallel to the local magnetic field. 
In the proton phase, space accelerated particles are visible upstream and downstream. A large increase in the proton temperature can be inferred from the width of the proton distribution.
For the test-particle electrons, the initial Maxwellian flux in the far upstream widens in the precursor (between $x=1000\,c/\omegap$ and $x=2000\,c/\omegap$), indicating an increase in temperature.
Furthermore, a population of counter-propagating particles with positive $v_\parallel$ is present upstream. This indicates a reflection due to magnetic mirroring near the shock transition. The inset shows a region close to the shock transition, where this effect is more pronounced.
Upon a shock crossing, the width of the distribution in velocity changes only slightly and only a minor increase in the temperature of the test-particle electron population is expected. 

\begin{figure}[ht]
   \includegraphics[width=\linewidth]{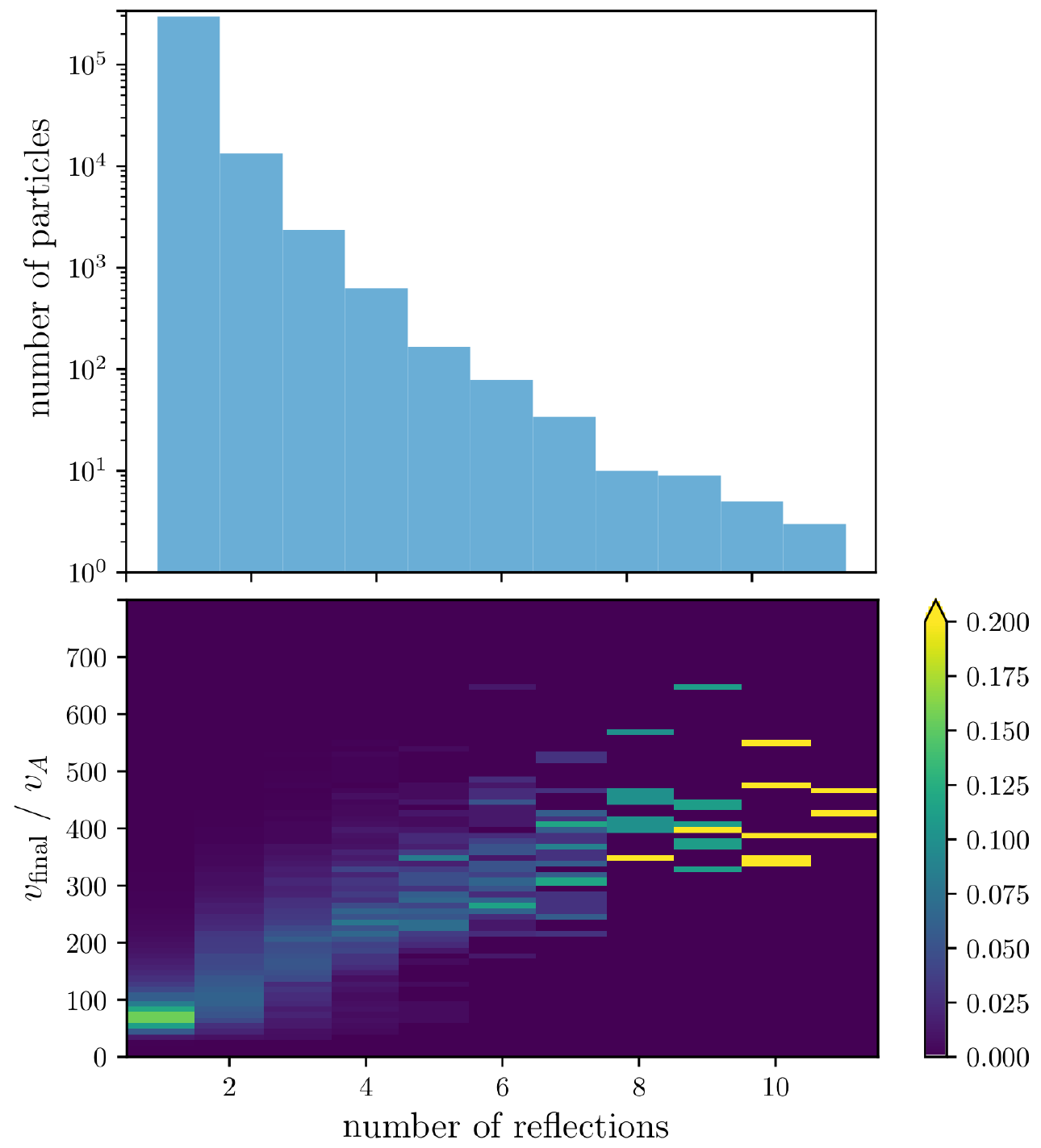}
        \caption{(top) -- Reflection on the shock histogram in the time interval $275 < t\,\omegac < 300.$ (bottom) -- Velocity of the reflected electrons after the last reflection in the time interval $275 < t\,\omegac < 300$  for a simulation (T) with $\ds v_0/\vA=10.$ The color denotes the fraction of reflected test electrons.}
        \label{fig:number_reflection}
\end{figure}

\begin{figure}[ht]
    \includegraphics[width=\linewidth]{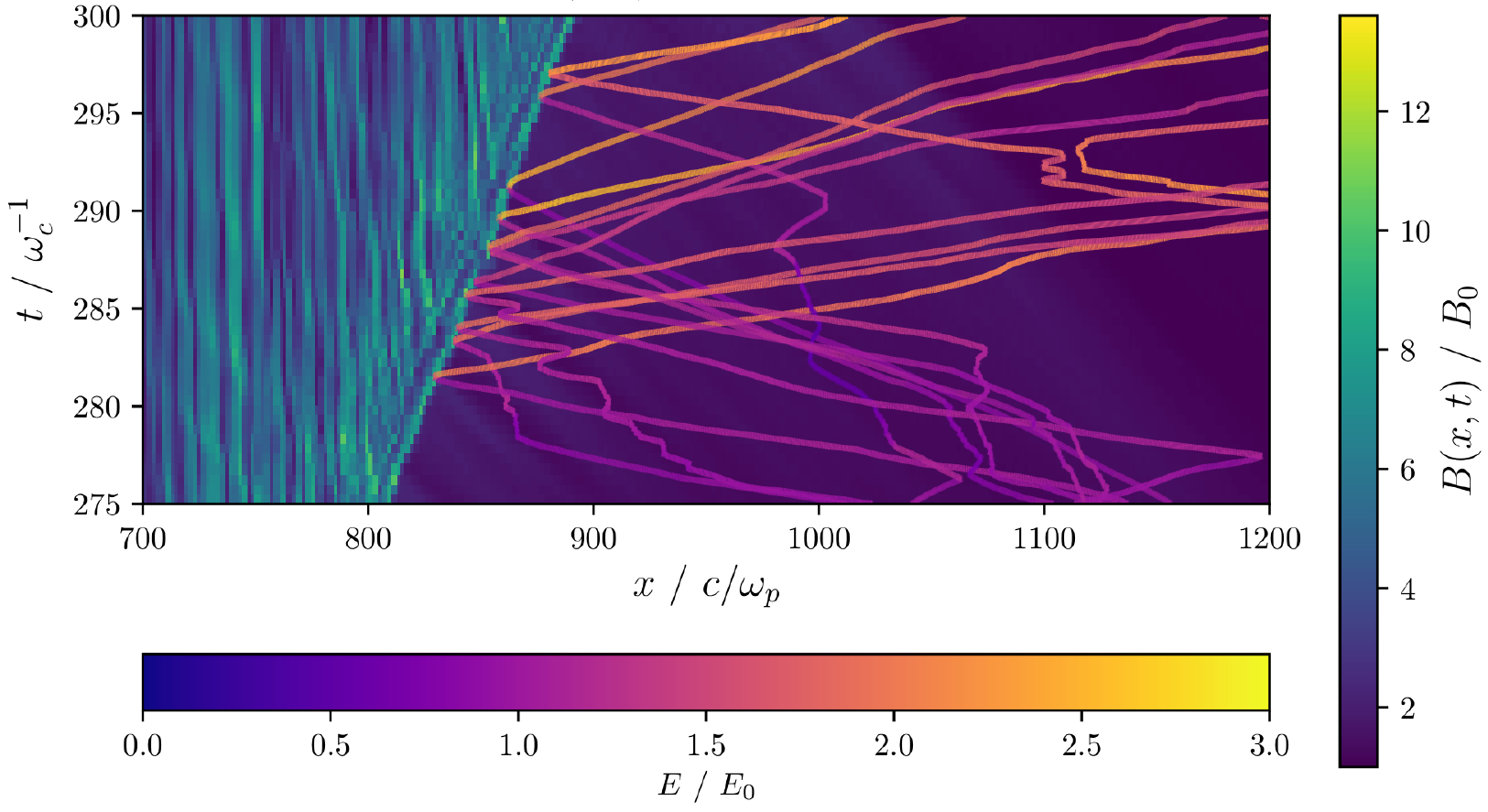}
        \caption{Trajectories of the electrons that are reflected off of the shock in the magnetic field $\vert B(x,t)\vert$ for a simulation (T) with $\ds v_0/\vA=10.$  
        The line color denotes the energy of the test-particle electrons in terms of their initial energy at the start of the tracing at $t\omegac=275.$}
        \label{fig:e_reflection}
\end{figure}

The multiple reflection of electrons from the shock may
significantly increase their energy.
The number of these %\st{types of} 
electrons, experiencing repeated reflections at the shock front and getting scattered by the upstream ion-generated turbulence, is considerable, as suggested 
by the histogram of the  test-electron reflection in the time interval $275 < t\,\omegac < 300,$ which is shown in Figure~\ref{fig:number_reflection} (top frame).
With an increasing number of reflections, the velocity of the test particles increases on average as well; this can be seen in Fig.~\ref{fig:number_reflection} (bottom frame).
We have traced some reflected particles 
and plotted the results in Fig.~\ref{fig:e_reflection}. 
The background shows the amplitude of the  magnetic field $\vert B(x,t)\vert$. The line color denotes the energy of the traced electrons in terms of their  initial 
energy at $\ds t\,\omegac=275$. It is clearly visible that all the traced test particles gain energy upon reflection and in the interaction with the proton-driven turbulence in the shock precursor\footnote{The idea of the electron acceleration by mirror reflection and trapping in the ion-scale turbulence in the quasi-parallel shock was put forward two decades ago by G. Mann \& H.-T. Cla{\ss}en \citep[see e.g.,][]{Mann_Classen1995,Classen_Mann1997}. They included test-particle modeling and proposed scenarios for observations in the solar context.}.
This can also be inferred from the phase space plot, shown in Fig.~\ref{fig:phase_space}, where particles with large positive $v_\parallel$
are present.  This process might be important for the injection of electrons into the DSA, though the downstream energy spectrum of the test-particle electrons does not show a clear power-law tail. 
We note that injection is not necessarily equivalent to the full DSA acceleration and the goal of our paper is to show the possibility of 
the first, most critical part of the two. 
Moreover, a clear power-law tail is not a prerequisite for injection. In the DSA context, the electron injection is understood to be an emergence of an electron population that can potentially be scattered by the ion-generated waves.
\begin{figure*}[h]
    \includegraphics[width=0.35\linewidth]{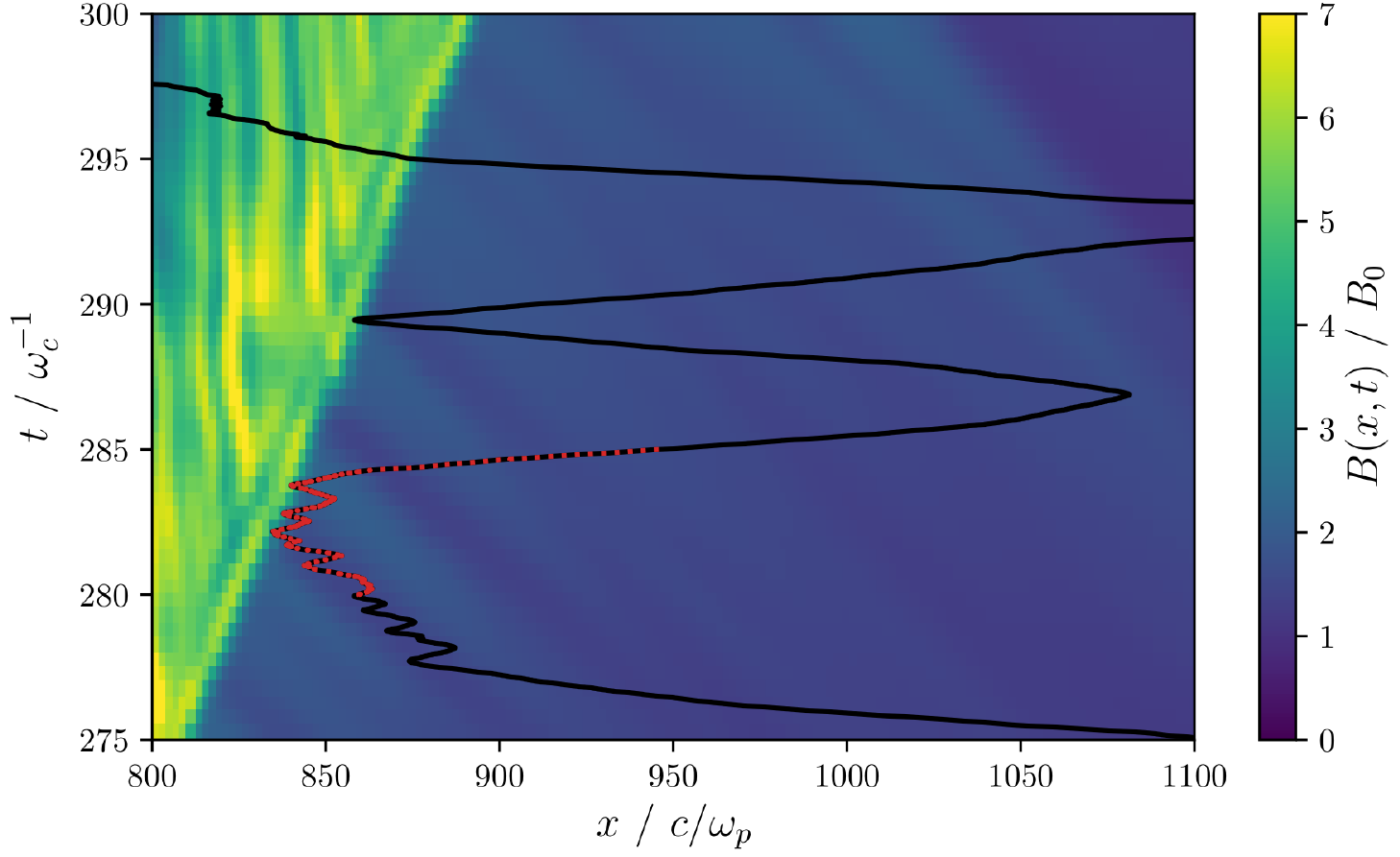}
   \includegraphics[width=0.319\linewidth]{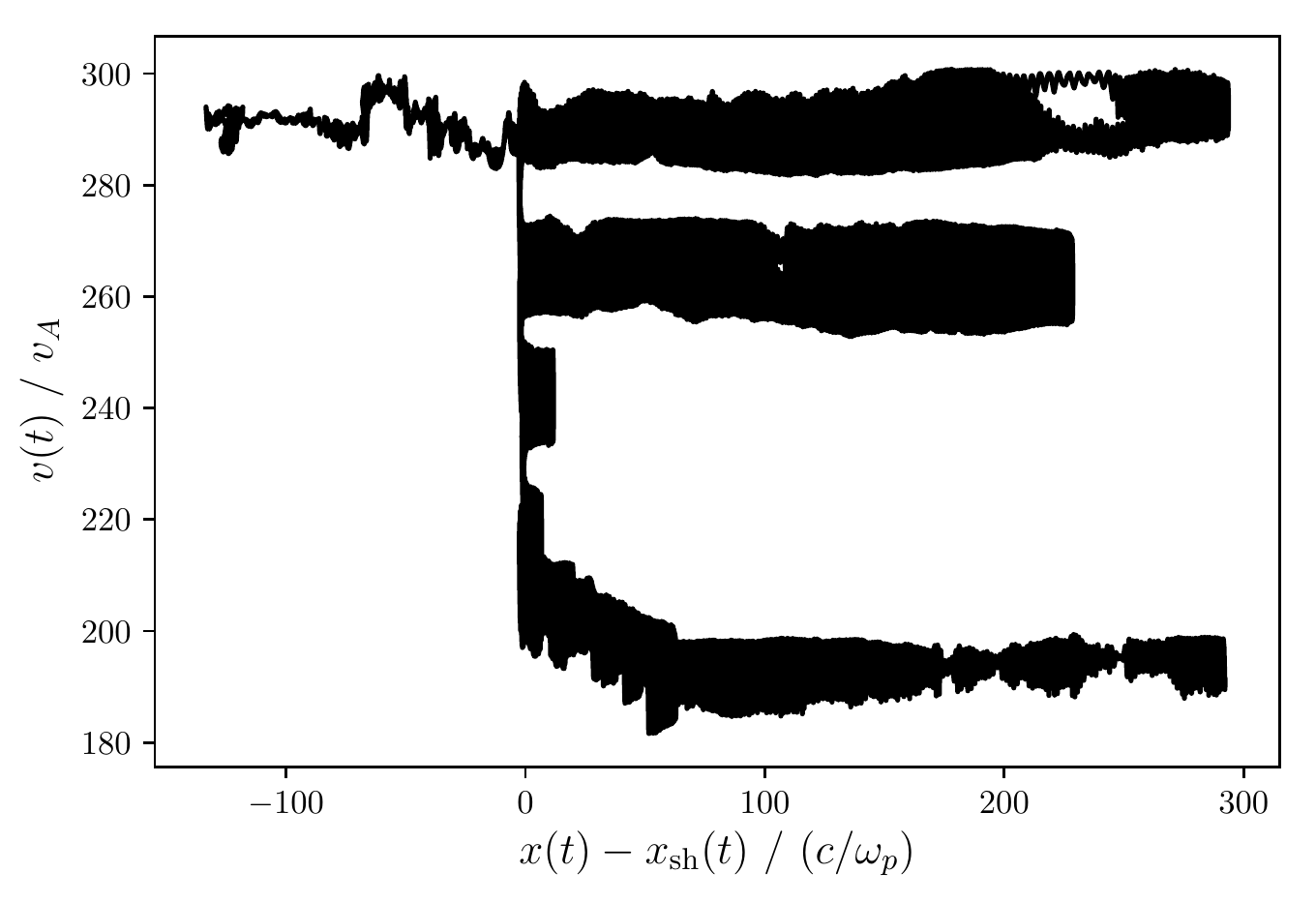}
   \includegraphics[width=0.319\linewidth]{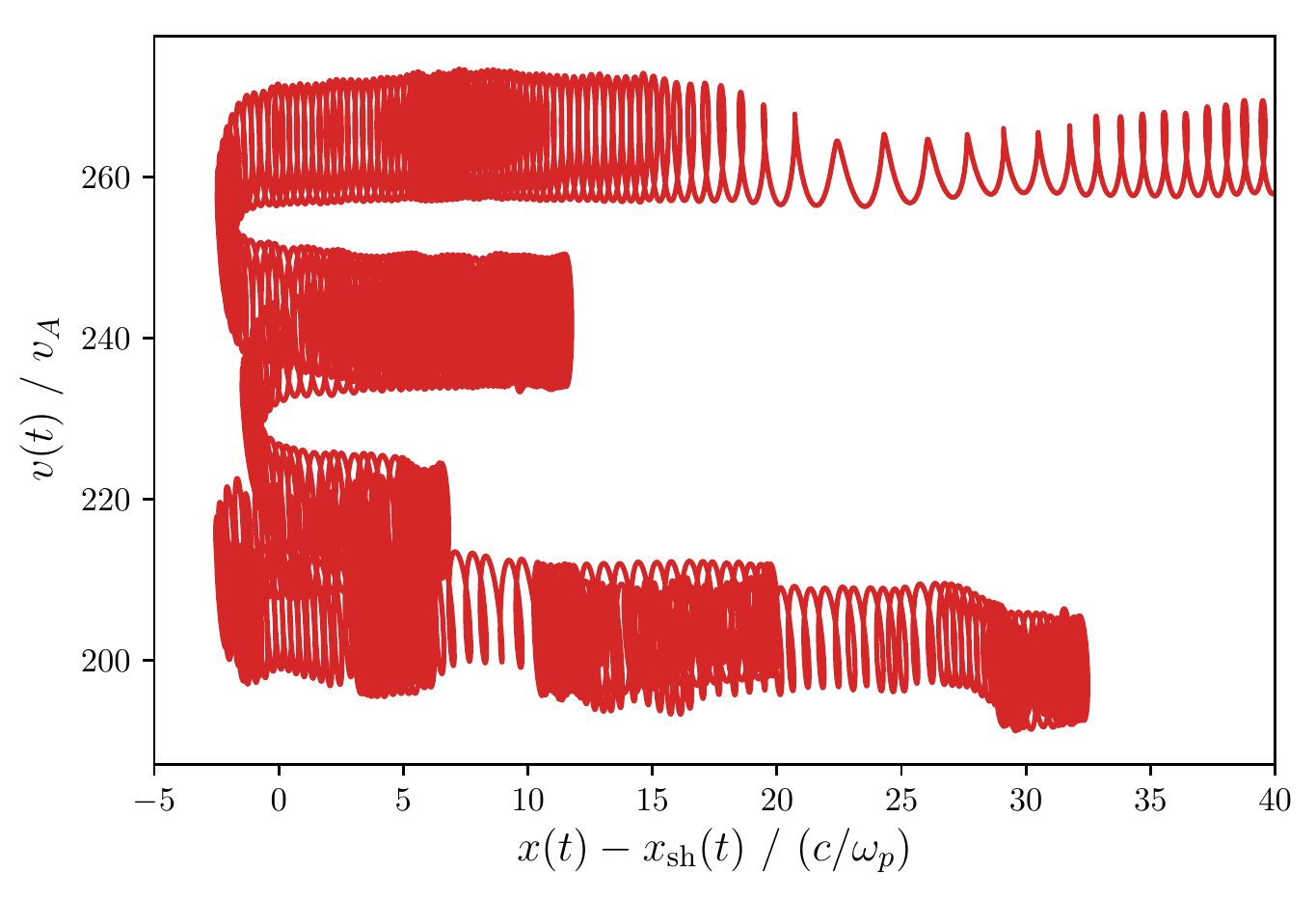}
        \caption{Trajectory of an electron in the $(x,t)$ and $(x-x_\mathrm{sh},v)$ spaces in the time interval $\ds 275 < t\,\omegac < 300$ for a simulation (T) with $v_0/\vA=10.$ The background (left frame) shows the amplitude of the  magnetic field $\vert B(x,t)\vert$. The test-electron thermal velocity that is far upstream is $\ds V_{\Te}/\vA=\sqrt{m_\mathrm{p}/\me} = 20.$ The relatively high velocity of the traced particle ($\ds v/\vA\sim 200$) at $t\,\omegac=275$ indicates that it has already been energized in the precursor and possibly during encounters with the shock at earlier times. }
        \label{fig:individual}
\end{figure*}
In Fig.~\ref{fig:individual}, we follow an exemplary test electron which, after being reflected off of the shock because of magnetic mirroring, remains trapped for a while upstream, close to the  shock front. The relatively high velocity of this electron at $t\omegac=275$ indicates that it has already been energized in the proton-driven turbulence developed in the precursor.
The results of a recent study by \cite{Guo2015}, where the test-particle electrons were propagated  in a predefined turbulent electromagnetic field in the shock region, 
suggest that electrons return back to the shock due to mirroring reflection off of the large amplitude waves upstream. The exemplary trajectory, which is shown in  Fig.~\ref{fig:individual}, indicates that there is indeed a mirroring process upstream. 
The wave spectra upstream of the shock transition, seen in Fig.~\ref{fig:spectra-5}, show that strong forward propagating waves, with the group and phase velocities that are higher than the Alfv\'en velocity with $\ds 0.05<k_x\cdot (c/\omegap)<0.2$ and $0.2<\omega/\omegac< 1$, are excited so that the resonance condition $\omega - k_x v_x = 0$ can be fulfilled for the electrons having $v_x \sim 20 \vA.$ 
\begin{figure}[ht]
   \includegraphics[width=1\linewidth]{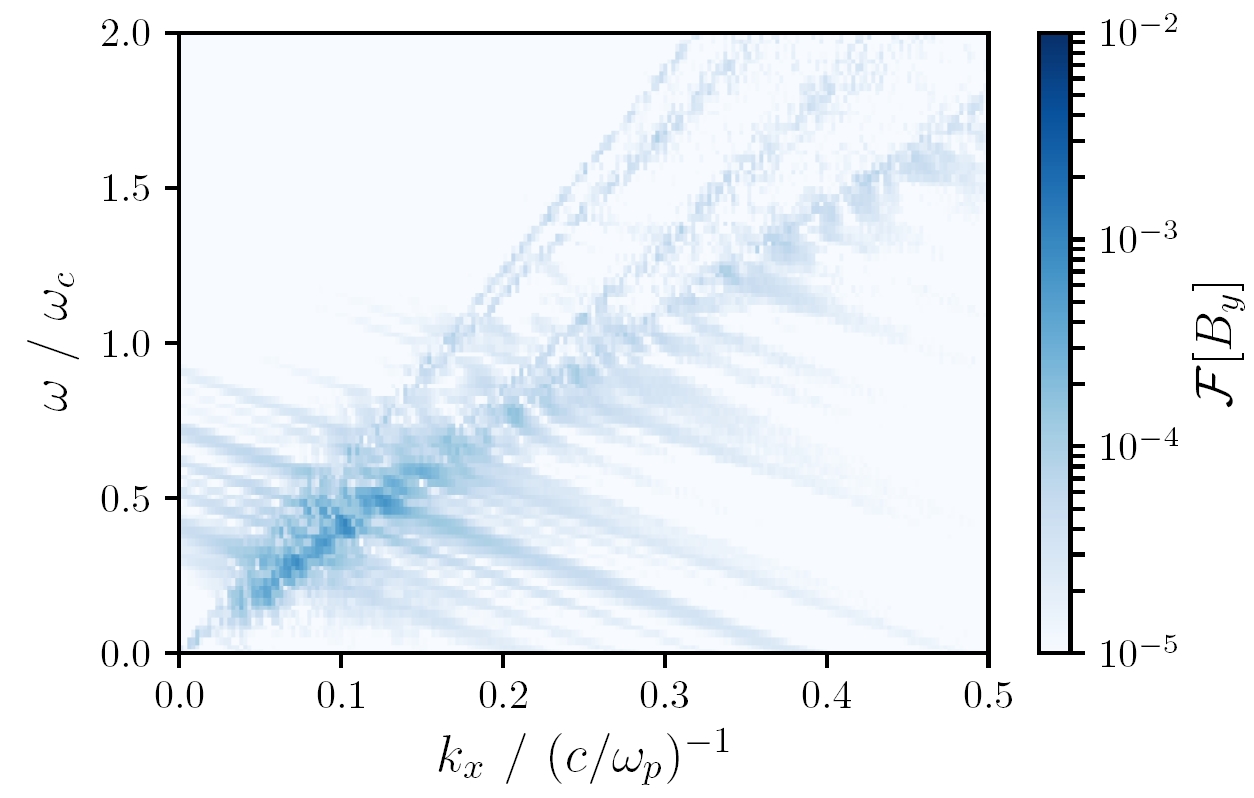}
        \caption{Spectrum of the magnetic field upstream for the simulation with $v_0/\vA=5.$ }
        \label{fig:spectra-5}
\end{figure}

\begin{figure}[ht]
    \centering
\includegraphics[width=\linewidth]{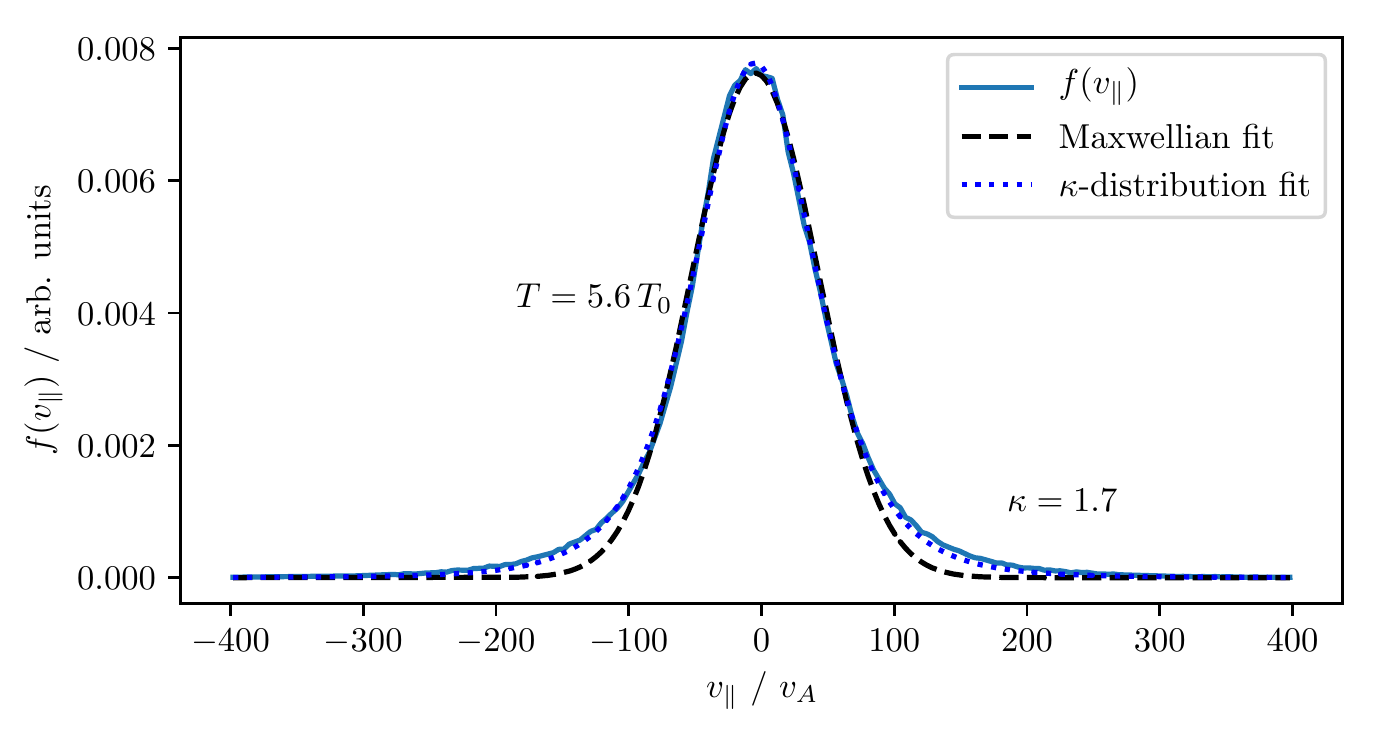}
    \caption{Velocity distribution of $\ds f_\mathrm{e}(v_\parallel)$ for the test-particle electron population 
    downstream of the shock transition at $t\,\omegac=400$ for a simulation (T) with $v_0/\vA = 10.$ The center of the distribution is described by a Maxwellian well (dashed-line). A suprathermal tail is clearly visible. 
    The relative amplitude of the tail at $v=150\,\vA$ is significantly higher, approximately
5500 times, than the value obtained 
    for the Maxwellian core electron distribution (electron fluid).}
    %$\approx 0.05,$ whereas for the Maxwellian core electron distribution (electron fluid), the value $9\cdot 10^{-6}$ is obtained.}
        \label{fig:velocity_dist}
    \end{figure}

The downstream distribution $\ds f_\mathrm{e}(v_\parallel)$ 
at $\ds t\,\omegac=400$ for a simulation (T) with $\ds v_0/\vA = 10$ is plotted in Fig.~\ref{fig:velocity_dist}.  This is a low-density addition to the electron core distribution on which the hybrid simulation operates. Since both the high temperature part, which is described by a Maxwellian\footnote{The main part of the downstream test-electron velocity distribution is properly described by a Maxwellian for all simulations (T). In the (B) and (S) cases with $\v0e/\vA=100$ and $\MA=6-10$, the complete electron thermalization downstream is not achieved and the velocity distribution there is "shell"-like, see Fig.~\ref{fig:shell}.} well (dashed line), and the energetic tail are absent in the core, this correction, however small in density, is physically important.  We emphasize that the 
%\LEt{ Please avoid the use of the slash. For details, refer to Sect. 3.6 of the A\&A Author's Guide.}
tail-to-core content in the test-particle electron population is vastly different from that 
of the fluid electrons. Behind the shock transition, the relative amplitude $\ds f^\mathrm{TP}(150\,\vA)/f^\mathrm{TP}(0)$ of the tail of the test-electron distribution at $v=150\,\vA$ is approximately 0.05; whereas, for the core electron distribution, it takes the value $\ds f^\mathrm{fluid}(150\,\vA)/f^\mathrm{fluid}(0)\simeq 9\cdot 10^{-6}$.
We note that the velocity distribution of the suprathermal tail population  fits with the Kappa ($\kappa$-) distribution best 
\[
f_\kappa(v) \propto (\kappa w_\kappa^2)^{-\frac{3}{2}}\frac{\Gamma(\kappa+1)}{\Gamma(\kappa-{1}/{2})}\left(1+\frac{v^2}{\kappa w_\kappa^2}\right)^{-(\kappa+1),}
\]
which is shown by a dotted line  in Fig.~\ref{fig:velocity_dist}.
Here, $\ds w_\kappa=\sqrt{(2\kappa-3)k_\mathrm{B} T/\kappa m}$ 
and $\Gamma$ is the Gamma function. 
Kappa distributions are frequently employed to describe the velocity distribution of collisionless plasmas out of thermal equilibrium \citep{Lazar-AA2016}. These include 
space and astrophysical plasmas from solar wind  and planetary magnetospheres to the heliosheath and beyond to interstellar and intergalactic plasmas \citep[see e.g.,][and references thererein]{Feldman1983,Pierrard2010,Oka2018,Kappa_Solar2018,Wilson_partition2019,livadiotis2017kappa}.

\begin{figure}[ht]
    \includegraphics[width=\linewidth]{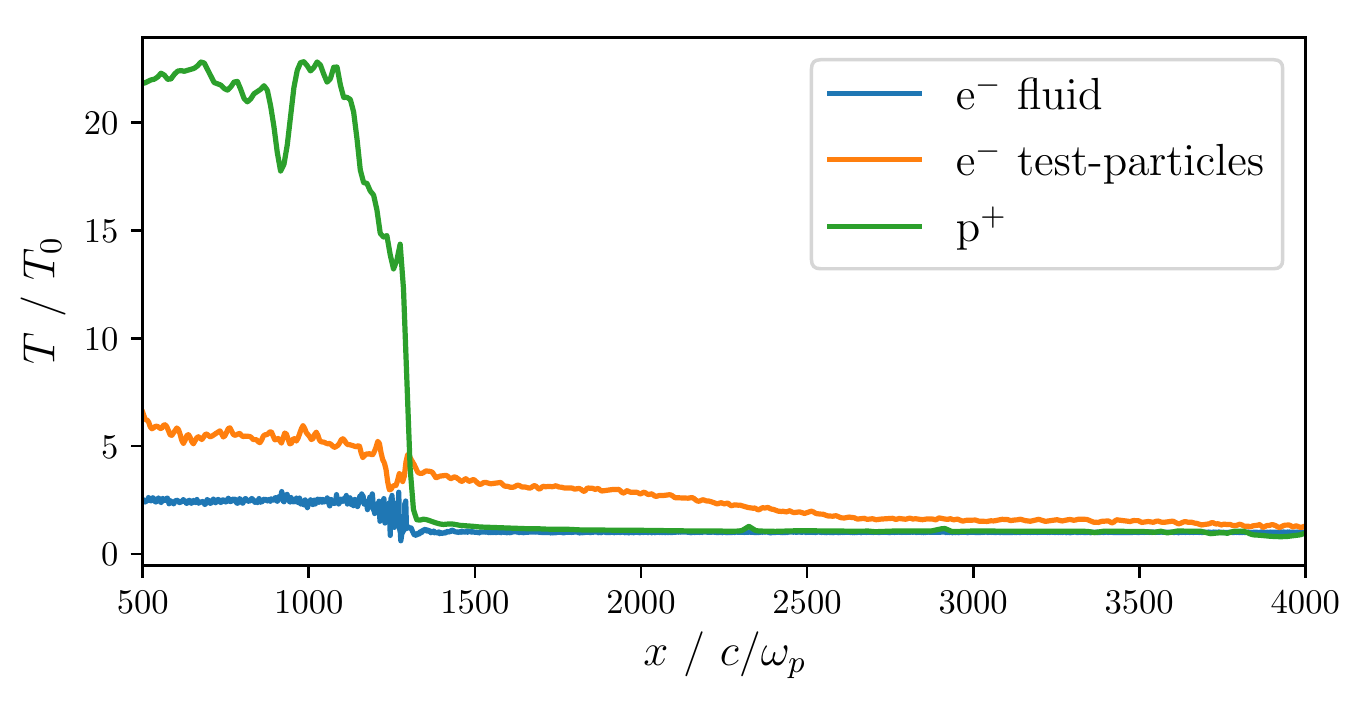}
        \caption{Spatial dependence of the temperature for the electron fluid (blue), test-particle 
                         electrons (orange), and protons (green) at $t\,\omegac=400$ for a 
                         simulation (T) with $\ds v_0/\vA = 10.$ Independent of the initial test-electron distribution, (T), (B), or (S), their temperature space profiles show significant heating in the precursor.}
        \label{fig:Temp_profile}
\end{figure}

\begin{figure}[ht]
       \includegraphics[width=\linewidth]{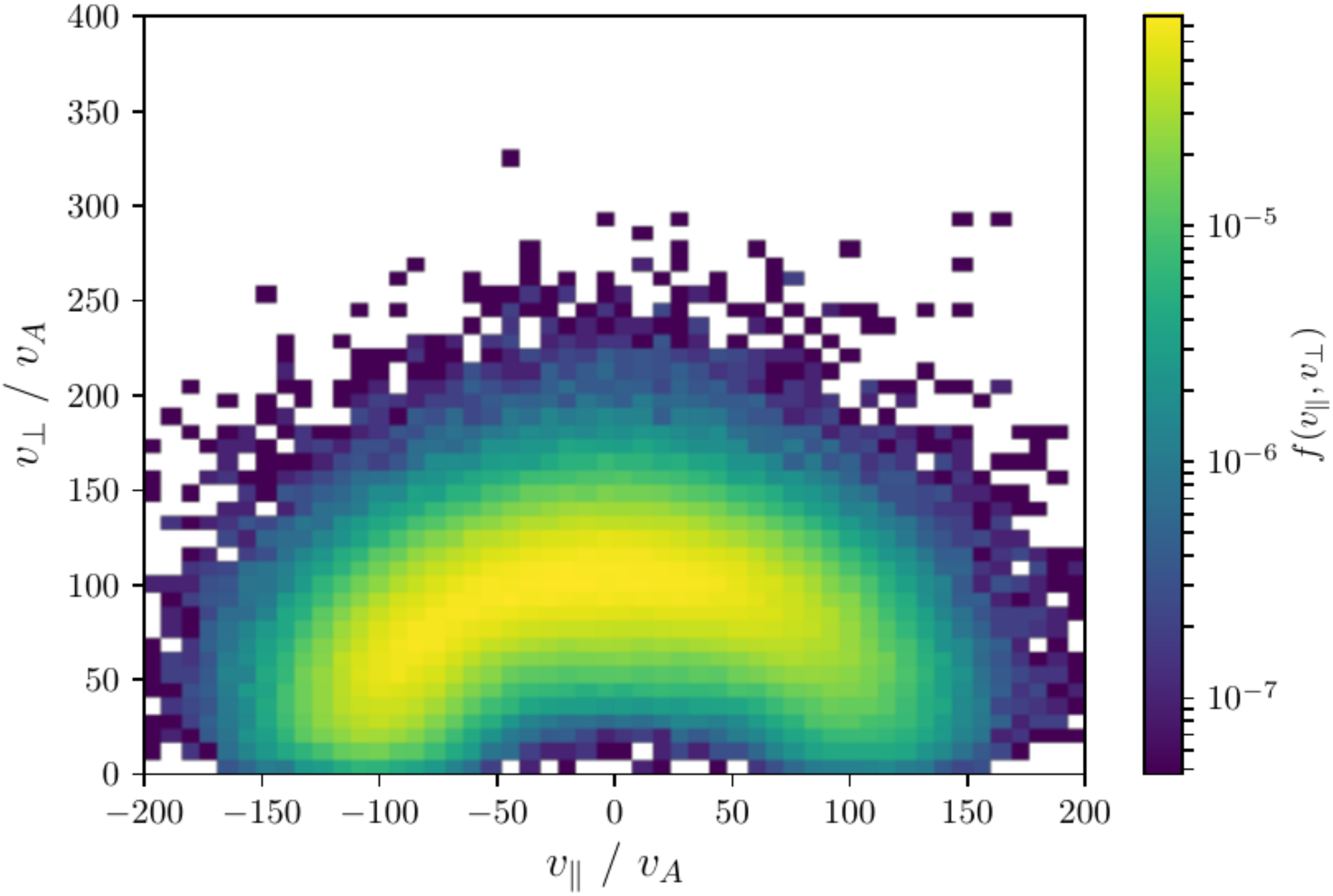}
        \caption{Downstream velocity distribution of the test electrons $f(v_\parallel,v_\perp)$  along 
        field line for a simulation (B) with $\ds v_0/\vA = 5$ averaged over the whole downstream region.}
        \label{fig:shell}
\end{figure}

The spatially dependent temperature profiles of protons (green), test-particle electrons (oran\-ge), and electron fluid (blue) are plotted in Fig.~\ref{fig:Temp_profile}  for a simulation (T) with an upstream flow velocity of 
$v_0/\vA = 10.$ It is apparent that the temperature profile of the test particles differs from the temperature of the electron fluid, which in our hybrid model have to follow $\ds \Te(x)/T_0 = \left(\ne(x)/n_0 \right)^{\gamma-1}$ because of the adiabatic closure between the density and pressure.
The downstream temperature of the test-particle electrons is approximately two times higher than the fluid temperature.  A gradual increase in the temperature of the test particles toward the shock transition in the region between $\ds x=1200\,c/\omegap$ and $\ds x=2600\,c/\omegap$ points to a considerable potential of the precursor wave field, which is supplied by ions, to preheat the electrons before they are shocked at the subshock\footnote{Though our simulations clearly indicate that most of the electron heating occurs in the precursor, there might be more heating at the ramp as well. We do not see this additional heating because our simulation is limited.}. Independent of the initial test-electron mean energy (distributions (T), (B), or (S)), the obtained temperature space profiles confirm the heating of the test electrons in the precursor.
We emphasize here 
that this considerably heated test-particle electron population,  by  definition,  does  not  generate electric or magnetic fields, nor exerts additional pressure on the background plasma. 

To obtain the dependence of the suprathermal (with respect to the core-electron population) electron temperature on the shock Mach number, $\MA = v_\mathrm{sh}/\vA$, we performed simulations for a range of initial flow velocities, $v_0$ 
(i.e., different shock velocities, $v_\mathrm{sh}$), 
and we calculated the downstream  test-electron temperatures for different initial test-electron distributions (T), (B), and (S).
The resulting proton, $\Ti,$  and test-electron temperatures, $T^\mathrm{h}_\mathrm{e},$ in terms of the far-upstream plasma temperature, $T_0,$ are summarized in the Table~\ref{tab:temp_data}. We note that the downstream electron velocity distribution for all the simulations (T) 
and for the simulations (B) and (S) and $v_0/\vA \ge 10$, 
is properly described by a Maxwellian, see Fig.~\ref{fig:velocity_dist}, and the temperature is extracted from the Maxwellian fit. For simulations (B) -- "beam" and (S) -- "shell" and relatively low shock Mach numbers ($v_0/\vA=5,7$ corresponding to $\MA=6.9, 9.6$), the downstream test-electron distribution function is "shell-like", as demonstrated in Fig.~\ref{fig:shell}, so that the "thermal" velocity of the test electrons (indicated by a $^*$ symbol in Tab.~\ref{tab:temp_data}) is approximately equal to the radius of the downstream velocity "shell".
Additionally, we calculated the velocity distribution for test electrons that reside downstream for a certain time: $t_\mathrm{ds}\omegac=175$ for $v_0/\vA=5$ and $t_\mathrm{ds}\omegac=100$ for $v_0/\vA=7$. These electrons 
had time to thermalize, and their distribution is close to a Maxwellian. The numbers in brackets in the Table~\ref{tab:temp_data} refer to the temperatures that were obtained accordingly.

\begin{table}[h]
\caption{Downstream proton $\Ti$ and 
test electron $T^\mathrm{h}_\mathrm{e}$ temperatures.}
  \centering
\begin{tabular}{c|c|c|c|c|c|c}
\hline
\hline
\rule{0pt}{4ex} 
$\ds\frac{v_0}{\vA}$ & $\MA$ & $\ds\frac{\Ti}{T_0}$ &\multicolumn{3}{c|}{$T^\mathrm{h}_\mathrm{e}/T_0$}& $\v0e/\vA$\\ 
    &&& (T) & (B) &(S) & (B)\\
\hline
\rule{0pt}{4ex}
    5   & 6.9   & 6.5   & 3.5   & 6.5$^*$               &                   & 30\\
        &       &       &       & 27.3$^*$ (6.4)        & 27.3$^*$ (7.3)    & 100\\
    7   & 9.6   & 11.2  & 3.7   & 4.1                   &                   & 30\\
        &       &       &       & 13.1$^*$ (6.5)        & 25.3$^*$ (12.4)   & 100\\
    10  & 13.3  & 20.7  & 6.2   & 5.9                   &                   &  30\\
        &       &       &       & 18.8                  & 17.4              & 100\\
    15  & 19.6  & 44.6  & 7.6   & 14.5                  &                   & 30\\
        &       &       &       & 29.2                  & 28.0              & 100\\
        \hline
    \end{tabular}
    \label{tab:temp_data}
\end{table}

\begin{figure}[t]
\rightline{
    \includegraphics[width=0.95\linewidth]{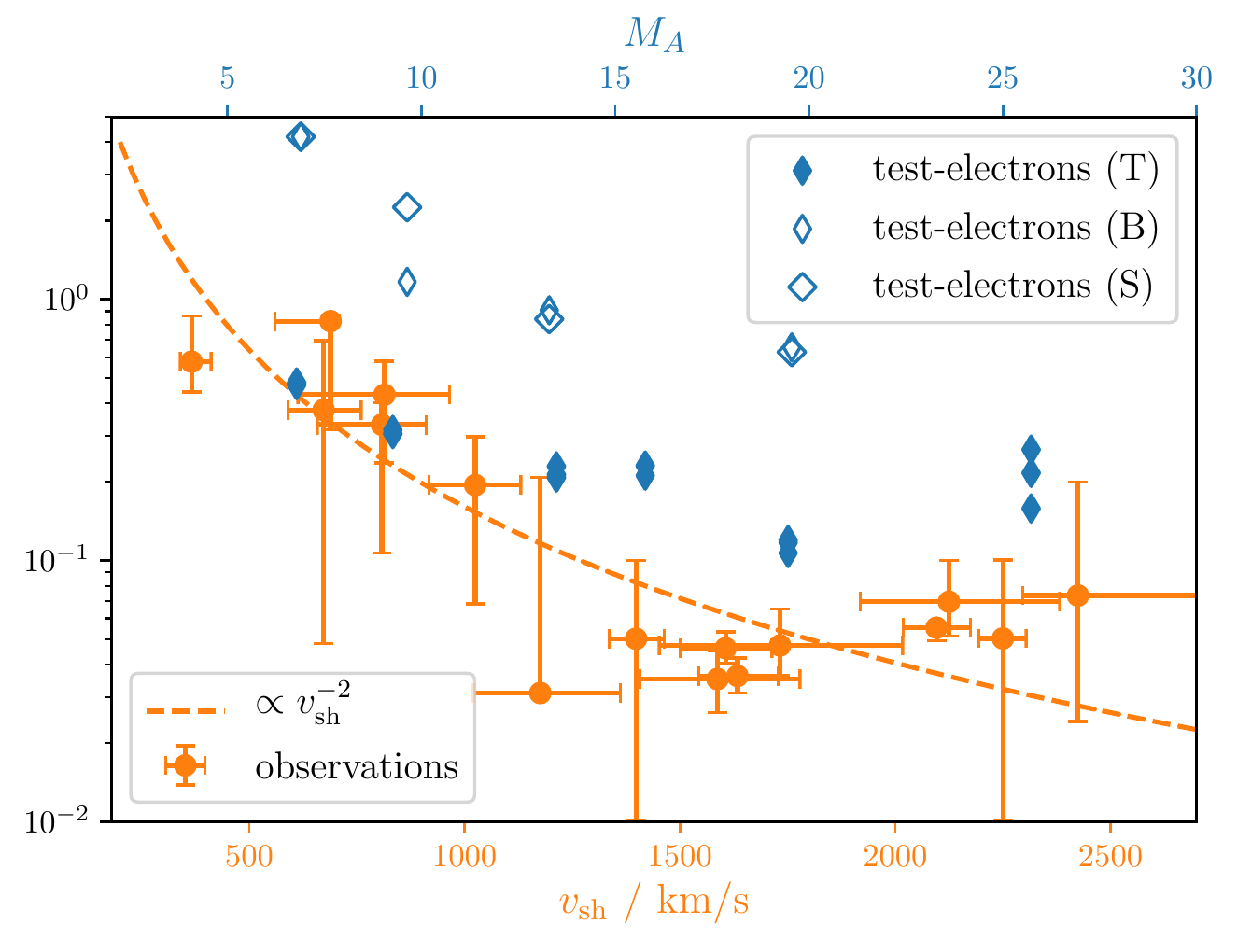}
}
        \caption{Test-electron-to-proton temperature ratio $\left(T^\mathrm{h}_\mathrm{e}/\Ti\right)(\MA)$ extracted from the simulations (for the upstream distribution (T) - blue filled diamonds - three time moments for every $\MA$ are shown) together with the temperature ratios, $\left(\Te/\Ti\right)(v_\mathrm{sh}),$ determined from observations of Balmer-dominated
                         shocks \citep{Ghavamian2013, Ghavamian2007,vanAdelsberg2008} (orange).}
        \label{fig:TeTp}
        
\end{figure}

The results, which are displayed in Fig.~\ref{fig:TeTp} with blue diamonds, provide evidence
that the test-electron-to-proton temperature ratio, $T^\mathrm{h}_\mathrm{e}/\Ti$, is a decreasing function of the shock Mach number with a tendency for saturation at high $\MA.$ For comparison purposes, in the same graph, we show the electron-to-proton temperature ratios that were determined from observations of Balmer-dominated shocks (orange symbols) as a function of the shock velocity\footnote{The shock Mach numbers were not known because the environmental conditions of the SNR shocks could not be determined directly.} $v_\mathrm{sh}$ \citep{Ghavamian2013, Ghavamian2007,vanAdelsberg2008}. 
The values of $v_\mathrm{sh}$ and the temperature ratio $\Te/\Ti$ were extracted from the width of the broad and narrow components of the H$\alpha$ line profile (see e.g., Fig.~1 in \cite{Ghavamian2013}).
\noindent The observational data, as can be seen in Fig.~\ref{fig:TeTp}, show that $\Te/\Ti$ decreases with increasing shock velocity; additionally, at high $v_\mathrm{sh}$, it seems to saturate on a level which is higher than the mass proportional heating, expected from the Rankine-Hugoniot jump conditions $\ds k_\mathrm{B}\, T_\mathrm{e,i} = \frac{3}{16} m_\mathrm{e,i} \, v_\mathrm{sh}^2.$  
For an analysis of the observations,  \cite{Ghavamian2007} applied a model of electron heating in which a constant level of electron heating over a wide range of shock velocities 
\citep[see also][]{Bykov1999}
is implied. It is also assumed that the ion heating increases with the shock speed.   
%\st{an increase in the shock speed ion heating is assumed.}
A scaling $\Te/\Ti \sim v_\mathrm{sh}^{-2}$ was found to fit to the observational data best \citep{Ghavamian2013} (orange-dashed line in Fig.~\ref{fig:TeTp}).
Instead, a function $\ds T^\mathrm{h}_\mathrm{e}/\Ti \sim M^{-1}$ is best fitted to our simulation results for the suprathermal electrons.  
A relatively high Alfv\'en velocity of $\vA = 90~$km/s has to be assumed to make a comparison of the measurements with the simulation data. This is about four times larger
than the velocity one would expect when using the standard parameters of the interstellar medium (for $B=3 \,\mu \textrm{G}$ and $n\!=\! 0.1 \,\textrm{cm}^{-3}$, the Alfv\'en velocity equals $\vA \simeq 20~$km/s). 
However, SNR environments are diverse and if, in addition,  the large-scale field is amplified, one might have to consider the Alfv\'en velocity 
in the amplified field, 
which can be as high as $\delta B/B = 4-10$.

\section{Discussion and conclusions}
Our simulations indicate that an equilibration of test electron and ion temperatures  does not occur. 
A decrease in the shock Mach number electron-to-ion temperature ratio, $\left(T^\mathrm{h}_\mathrm{e}/\Ti\right)(\MA),$  is observed instead.  We note that our test electrons correspond to the suprathermal, with respect to the core-electron population, which is part of the distribution function and the simulations do not tell how abundant this electron population might be. The shock velocity dependence of the ratios of the core electron-to-proton temperature, $\left(\Te/\Ti\right)(v_\mathrm{sh})$, that were experimentally determined for Balmer-dominated shocks has a similar trend. For strong shocks, the observational data suggest a saturation or even an upturn of the temperature ratio. 
The observed scaling, $\Te/\Ti(v_\mathrm{sh})\propto v_\mathrm{sh}^{-2},$ is supported by theoretical predictions \citep{Vink2015} that a dependence of $\Te/\Ti \sim M_\mathrm{s}^{-2}$, where $M_\mathrm{s} = v_\mathrm{sh}/c_\mathrm{s}$ is the sonic Mach number, can be obtained by solving the Rankine-Hugoniot 
jump conditions with the assumption that the enthalpy flux is conserved for each particle species separately. 
This yields an $\sim M_\mathrm{s}^{-2}$ behavior for shocks with Mach numbers in the range 
\begin{equation}
  \sqrt{ \frac{2}{\gamma-1} \, \frac{\mu}{\mi} \, \frac{r^2}{r^2-1} } 
  < M_\mathrm{s} <
  \sqrt{ \frac{2}{\gamma-1} \, \frac{\mu}{\me} \, \frac{r^2}{r^2-1} },
  \label{eq:vink}
\end{equation}
\citep[see Eq.~(14) in][]{Vink2015} with $\mu = (\mi+\me)/2$ being the average mass and $r$ being the shock compression ratio. 
For the increased electron-to-ion mass ratio and 
parameters used in our simulations, Eq.~\eqref{eq:vink} translates to $1.4 < \MA < 23$.
In the case of efficient CR acceleration, that is, when considering the CR pressure in the precursor, \citet{Vink2015} also predict a higher level of saturation of the temperature ratio toward a high $M_\mathrm{s}$\footnote{\citet{Ghavamian2007} attributed  
the observed scaling to the electron heating in a shock precursor.  
\citet{Laming_2014} discuss the implication of the shock velocity dependent precursor length on the electron heating and speculate that for shocks of higher velocities than those considered in %\LEt{ punctuation issue.}
\citet{Ghavamian2007} and \citet{vanAdelsberg2008}, the electron temperature could even rise with increasing shock speed.}.  The predicted range for the $M_\mathrm{s}^{-2}$ scaling, with a dependence on $\me$ and $\mi$, disfavors fully kinetic simulations, 
where the reduced mass ratios of $m_\mathrm{p}/\me = 64 - 100$ are used regularly \citep{Park2015}. In this case,  $M_\mathrm{s}^{-2}$  scaling may only occur in a limited range of Mach numbers.  

To conclude, different models have been proposed to heat electrons in front of SNR shocks. All of these models have to rely on numerical simulations since the particle distributions in SNR shocks cannot be measured in situ. When focusing on nonrelativistic collisionless shocks, two scenarios have mainly been considered: heating due to lower hybrid waves in the precursor \citep{Laming2000} or a mechanism based on counterstreaming instabilities in front of the 
shock \citep{Cargill1988}. Both mechanisms work well for perpendicular shocks. In
%\LEt{ punctuation issue.} 
\citep{Rakowski2008} and \citep{Voelk1995}, it has been argued
that due to the amplification of the magnetic field ahead of the shock, the perpendicular component might be large enough for the models to also be applicable for quasi-parallel shocks. Our simulation shows that the well developed ion turbulence in the precursor is able to influence the dynamics of the electron population and is responsible for the electron preheating  in the quasi-parallel shocks as well.

\section*{Acknowledgments}
The research was supported by DFG grant 278305671.
% and in part by NASA ATP-program within grant 80NSSC17K0255. 
M.A.M. acknowledges the NASA ATP-program  support within grant
80NSSC17K0255, and the National Science Foundation under grant NSF PHY-1748958.
A.H. \& T.V.L. acknowledge the North-German Supercomputing Alliance (HLRN) for providing the computational resources for the simulations (project mvp00015). T.V.L. acknowledges the support by the state contract with ICMMG SB RAS (0315-2019-0009) and by the Ministry of Science and Higher Education of the Russian Federation (agreement 075-03-2020-223/3 within FSSF-2020-0018) in the part related to the analysis of the numerical results. 
\bibliographystyle{aa} 
\bibliography{Literature}

\end{document}